\documentclass[a4paper, 12pt]{article}
\usepackage{epsfig,amssymb,euscript,xspace, color}
\usepackage{amsmath,jheppub,mathtools,empheq,amsthm}
\usepackage{graphicx}
\usepackage{verbatim}
\usepackage{cancel}

\usepackage[T1]{fontenc} 
\usepackage{tikz,caption,subcaption,marvosym} 
\usetikzlibrary{decorations.markings,arrows,snakes}
\usepackage{epsfig,amssymb,euscript,xspace, color}
\usepackage{amsmath,mathtools,rotating}

\def\bea{\begin{eqnarray}}
\def\eea{\end{eqnarray}}
\def\be{\begin{equation}}
\def\ee{\end{equation}}



\begin{document}

\title{
{\large Chiral $\Lambda$-$\mathfrak{bms}_4$ symmetry of 3d conformal gravity}}
\author{Nishant Gupta$^{a,c}$ and Nemani V. Suryanarayana$^{b,c}$}
\affiliation{$^{a}$National Institute of Science Education and Research (NISER)\\
	Bhubaneswar 752050, Odisha,
	India \\
	\\
	 $^b$Institute of Mathematical Sciences,\\ Taramani, Chennai 600113, India \\
\\$^c$Homi Bhabha National Institute,  \\ Anushakti Nagar, Mumbai 400094, India }
\emailAdd{nishantg@niser.ac.in, nemani@imsc.res.in}
%\date{}

\abstract{
We propose mixed boundary conditions for 3d conformal gravity consistent with variational principle in its second-order formalism that admit the chiral $\Lambda$-$\mathfrak{bms}_4$ algebra as their asymptotic symmetry algebra. This algebra is one of the four chiral $\mathcal W$-algebra extensions of $\mathfrak{so}(2,3)$ and is a generalisation of the chiral $\mathfrak{bms}_4$ algebra responsible for soft theorems of graviton MHV amplitudes in ${\mathbb R}^{1,3}$ gravity to the case of non-zero negative cosmological constant. The corresponding charges are shown to be finite and integrable, and realise this non-linear ${\cal W}$-algebra.}
%We comment on the potential the role played by 3d conformal gravity in AdS$_4$/CFT$_3$ correspondence. }.

\maketitle

\section{Introduction}
It is well-known that the  $AdS_3$ gravity theories admit different asymptotic symmetry algebras (ASA) that are infinite dimensional extensions of $\mathfrak{so}(2,2)$ depending on the boundary conditions imposed. For example, those of \cite{Brown:1986nw, Troessaert:2013fma} give non-chiral 2d conformal algebras whereas those of \cite{Avery:2013dja, Apolo:2014tua} give a chiral extension of $\mathfrak{so}(2,2)$. The latter one is generated by one chiral stress tensor $T(z)$ with central charge $c$ and a triplet $J^a(z)$ (for $a = 0, \pm 1$) of currents ($h=1$) making an $\mathfrak{sl}(2,{\mathbb R})$ Kac-Moody algebra at level $\kappa$.\footnote{In general there is no condition on $c$ and $\kappa$ required by consistency, however, the one realised in \cite{Avery:2013dja} has $c=\kappa/6$.} Just as the algebra in \cite{Avery:2013dja} can be shown to be the only chiral extension of $\mathfrak{so}(2,2)$, there are exactly four chiral extensions of $\mathfrak{so}(2,3)$ algebra all of which turn out to be non-trivial ${\cal W}$-algebras \cite{Gupta:2023fmp}. One of them is the chiral $\Lambda$-$\mathfrak{bms}_4$ \cite{Romans:1990ta, Gupta:2022mdt},\footnote{The $\Lambda$-$\mathfrak{bms}_4$ algebra was introduced in the works  \cite{Compere:2019bua, Compere:2020lrt,Fiorucci:2020xto} and should be thought as the non-chiral extension of $\mathfrak{so}(2,3)$ .} the generalisation of the chiral $\mathfrak{bms}_4$ responsible for soft theorems of graviton MHV amplitudes in ${\mathbb R}^{1,3}$ gravity \cite{Banerjee:2020zlg, Banerjee:2021dlm} to the case of non-zero negative cosmological constant ($\Lambda$). The operator content of the chiral $\Lambda$-$\mathfrak{bms}_4$ algebra includes $(T(z), J^a(z))$ of  \cite{Avery:2013dja} along with a couple of chiral primary operators $G_s(z)$ (for $s= \pm 1/2$ with $h=3/2$) that are also current algebra primaries in the doublet representation of $\mathfrak{sl}(2, {\mathbb R})$. In terms of appropriately defined modes $\{L_m, J_{a,n}, G_{s,r}\}$ for $m \in \mathbb Z, r \in \mathbb{Z}+\frac{1}{2}$, the chiral $\Lambda$-$\mathfrak{bms}_4$  algebra has the following commutation relations\footnote{It was shown in \cite{Gupta:2022mdt} that this algebra emerges as the ASA of $AdS_4$ gravity with a well-defined set of chiral boundary conditions, generalising the corresponding realisation of chiral $\mathfrak{bms}_4$ algebra in $\mathbb{R}^{1,3}$ gravity \cite{Gupta:2021cwo}.}
\begin{align}
	\label{RESULT}
	[L_m, L_n] &= (m-n) \, L_{m+n} + \frac{c}{12} m (m^2-1) \delta_{m+n,0} \, ,\cr
	[L_m, J_{a,n}] &= -n \, J_{a, m+n}, ~~~ [L_m, G_{s,r}] = \frac{1}{2}(n-2r) \, G_{s,n+r} \,,\cr
	[J_{a,m}, J_{b,n}] &= -\frac{1}{2} \kappa \, m \, \eta_{ab} \, \delta_{m+n, 0} + {f_{ab}}^c \, J_{c,m+n}, ~~ [J_{a,n}, G_{s,r}] = G_{s',n+r} {(\lambda_a)^{s'}}_s \, ,\cr
	\!\!\!\!\! [G_{s,r}, \, G_{s',r'}] \!&= \!\epsilon_{ss'}  \Big[\alpha \Big(r^2 - \frac{1}{4} \Big)  \delta_{r+r',0} \! + \! \beta \,  L_{r+r'}  \! + \! \gamma \, (J^2)_{r+r'} \Big] \!\!+ \!\delta (r-r')  J_{a, r+r'}  {(\lambda^a)}_{ss'}  \cr & \cr
	{\rm with} ~~c = &  - \frac{6 \kappa \, (1+ 2 \kappa)}{5+2\kappa},~ \alpha = - \frac{1}{4} \gamma \kappa (3+2\kappa), ~ \beta =  \frac{1}{4}\gamma(5+2\kappa), ~ \delta = -\frac{1}{2} \gamma (3+2\kappa)\, ,\cr &
\end{align} 
Here $\eta_{ab} = (3a^2-1) \delta_{a+b,0}$, ${f_{ab}}^c = (a-b) \delta^c_{a+b}$, $(\lambda^a)_{ss'} = \frac{1}{2} \delta^a_{s+s'}$, $(\lambda_a)_{ss'} = \eta_{ab} (\lambda^b)_{ss'}$, $\kappa \ne -5/2$, and $\gamma$ can be fixed to be any non-zero function of $\kappa$ by rescaling the $G_{s,r}$ appropriately. Finally $(J^2)_n$ are the modes of the normal ordered quasi-primary $\eta^{ab} (J_aJ_b)(z)$. 
%This algebra turned out to be closely related to the algebra derived by Romans \cite{Romans:1990ta} in a different context.

 Another of the four chiral ${\cal W}$-algebra extensions of $\mathfrak{so}(2,3)$ is the conformal $\mathfrak{bms}_3$ \cite{Gupta:2023fmp} whose semi-classical version was already known to arise naturally as the ASA of 3d conformal gravity \cite{Fuentealba:2020zkf}. Here we address the question of whether one can obtain the chiral $\Lambda$-$\mathfrak{bms}_4$ algebra as well from the 3d conformal gravity. 
 
Majority of works in the literature on ASA of 3d conformal gravity  \cite{Fuentealba:2020zkf,Afshar:2013bla, Bertin:2012qw}  have been carried out in the first-order formulation described by the Chern-Simons theory with $\mathfrak{so}(2,3)$ gauge algebra \cite{Deser:1981wh,Deser:1982vy,Horne:1988jf}. In its second order formulation the 3d conformal gravity is described by Chern-Simons gravity with the action \eqref{CSG1}. This action is usually added to the Einstein-Hilbert action so that the resultant theory describes a topologically massive gravity \cite{Deser:1981wh, Deser:1982vy, Li:2008dq, Carlip:2008jk, Skenderis:2009nt}. The second order formulation remains much less explored  in comparison to the first-order one. In \cite{Afshar:2011qw, Afshar:2011yh} the authors did study holography of Chern Simons-gravity theory where they proposed Dirichlet-type boundary conditions and showed the existence of two copies of Virasaro algebra along with one abelian Kac-Moody as the asymptotic symmetry algebra when the Weyl factor of the bulk metric is not kept fixed. However, the resultant ASA does not include $\mathfrak{so}(2,3)$ global symmetry. 
%The treatment of \cite{Afshar:2011qw} included the calculation of charges using first-order formalism. 

Another motivation to study conformal gravity theory defined by the action \eqref{CSG1} is the role it plays in the holography of AdS$_4$ gravity. The action \eqref{CSG1} becomes relevant when one considers AdS$_4$ gravity using the Neumann boundary condition where the holographic stress tensor $T_{ij}=0$, and one expects the boundary action to be an induced gravity \cite{Compere:2008us}. One can further add a bulk Pontryagin term to the action of AdS$_4$ gravity
\begin{align}
	S_{pontryagin}=k \int d^4 x\,\sqrt{-g}\epsilon^{\mu\nu\sigma\lambda}\,R^{\alpha}_{~\beta\mu\nu}\,R^{\beta}_{\alpha\sigma\nu} \label{pontryagin}
\end{align} 
with an arbitrary dimensionless coupling constant $k$. The Pontryagin term is a total derivative,  equivalent to  Chern-Simons Gravity action  (\ref{CSG1}) at the boundary. Therefore, the variation of the action after adding all the boundary terms, necessary counter terms and  the Pontryagin term \eqref{pontryagin} will be given by
\begin{align}
	\delta S_{total}\sim \int d^3x\, \sqrt{g^{(0)}} \, (T^{ij}+k\, C^{ij})\,\delta g^{(0)}_{ij}
\end{align}
where $g^{(0)}_{ij}$ is the induced metric at the boundary. For Neumann  boundary conditions where $\delta g^{(0)}_{ij}\neq 0$, one sets the modified holographic stress tensor to zero. However, in the large $k$ limit, the term proportional to $T^{ij}$ can be neglected, and we can solve the variational principle by setting $C^{ij}=0$, which means we look for only those solutions whose boundary metric is conformally flat. Furthermore, if one calculates boundary effective action then the action \eqref{CSG1} should dominate the induced gravity action in the large-$k$ limit at the boundary and thus becomes relevant for AdS$_4$ holography.

In this paper, we analyse the asymptotic symmetries of the Chern-Simons gravity theory \eqref{CSG1} by considering mixed boundary conditions for the metric such that the solutions satisfy well defined variational principle. We further derive charges associated with residual Weyl transformation and diffeomorphisms using the modified covariant phase space formalism proposed by Tachikawa \cite{Tachikawa:2006sz}. We show that the charges are finite and integrable, and the algebra obeyed by them is the semi-classical (large-$\kappa$) limit of \eqref{RESULT}, the chiral $\Lambda$-$\mathfrak{bms}_4$ algebra \cite{Gupta:2022mdt, Romans:1990ta}. Our boundary conditions do not break the global $\mathfrak{so}(2,3)$ symmetries of 3d conformal gravity similar to \cite{Fuentealba:2020zkf} but unlike those of \cite{Afshar:2011qw, Afshar:2013bla, Bertin:2012qw}. We also carry out  computations in the first-order formalism for completeness obtaining results consistent with the second-order ones.

The rest of the paper is organised as follows. In section \ref{second-order} we provide details of our main computation in the metric formalism of 3d conformal gravity. In section \ref{first-order} we provide some details of the same computation performed in the first-order formalism and show that the results are consistent with those of section \ref{second-order}. In section  \ref{discussion} we summarize our result and discuss its potential relevance to AdS$_4$/CFT$_3$. In Appendix \ref{rev_CSG} we provide a brief review of the relevant covariant phase space formalism we used to obtain charges in the main text. 

\section{Chiral $\Lambda$-$\mathfrak{bms}_4$ from 3d Chern-Simons Gravity}
\label{second-order}
\label{CGSintro}
In its second-order formulation the action of the 3d conformal gravity is given by
\begin{align}
	S_{CSG}=\frac{k}{2}\int d^3x\,\sqrt{-G}\,\epsilon^{\lambda\mu\nu}\,\left(\Gamma^{\rho}_{\lambda \sigma}\,\partial_{\mu}\Gamma^{\sigma}_{\rho\nu}+\frac{2}{3}\,\Gamma^{\rho}_{\lambda \sigma}\,\Gamma^{\sigma}_{\mu\tau}\,\Gamma^{\tau}_{\nu\rho}\right) \label{CSG1}
\end{align}
where $k$, the level of the Chern-Simons action, is a dimensionless parameter, $G$ is the determinant of the metric $G_{\mu\nu}$, and $\Gamma^\mu_{\nu\sigma}$ are the Christoffel symbols 
%(\textcolor{red}{with $\Gamma^\mu_\nu := \Gamma^\mu_{\nu\sigma}dx^\sigma$ being Levi-Civita connections}) 
associated with $G_{\mu\nu}$. The completely antisymmetric 3d Levi-Civita tensor $\epsilon^{\mu_1\mu_2\mu_3}$ is  
\begin{align}
	\epsilon_{\mu_1\mu_2\mu_3}=\sqrt{-G}\,	\hat{\epsilon}_{\mu_1\mu_2 \mu_3}\,,~~	\epsilon^{\mu_1\mu_2\mu_3}=\frac{1}{\sqrt{-G}\,}	\hat{\epsilon}^{\mu_1\mu_2\mu_3}
\end{align}
where $\hat{\epsilon}^{\mu_1\mu_2\mu_3}$ is a pseudo-tensor defined by
\begin{align}
	dx^{\mu_1}\wedge dx^{\mu_2}\wedge x^{\mu_3}=\hat{\epsilon}^{\mu_1 \mu_2 \mu_3}\,dx^{1}\wedge dx^{2}\wedge dx^{3}=\hat{\epsilon}^{\mu_1 \mu_2\mu_3}\, d^3x \, .
\end{align}
%that takes the value \textcolor{red}{$\hat{\epsilon}^{r+-}=1$}.
Therefore, the volume element on the 3d manifold can be written as $\sqrt{-G}\,\epsilon^{\mu_1\mu_2\mu_3} d^3x$.
%\begin{align}
%	dx^{\mu_1}\wedge dx^{\mu_2}\wedge dx^{\mu_3}=	\sqrt{-G}\,\epsilon^{\mu_1\mu_2\mu_3} d^3x = \hat{\epsilon}^{\mu_1 \mu_2\mu_3}\, d^3x.
%\end{align}
The equation of motion derived from the action \eqref{CSG1} is
\begin{align}
	C_{\mu\nu}= \epsilon_{\mu}^{~~\sigma\rho}\nabla_{\sigma}\left(R_{\rho\nu}-\frac{1}{4}R\,G_{\rho\nu}\right)=0 .\label{Cotten_tensor}
\end{align}
The tensor  $S_{\mu\nu}=R_{\mu\nu}-\frac{1}{4}\,R \, G_{\mu\nu}$ is the 3d Schouten tensor and $C_{\mu\nu}$ is the Cotton tensor. In 3d the equation \eqref{Cotten_tensor} is a sufficient and necessary condition for $G_{\mu\nu}$ to be conformally flat. Under Weyl transformations 
$$G_{\mu\nu} \rightarrow e^{-2\sigma} G_{\mu\nu}$$ 
the Cotton tensor  $C_{\mu\nu}$  is invariant (but this is not true in general for $d \geq 4$). The variations of the Lagrangian density in \eqref{CSG1} under diffeomorphisms and Weyl transformations are total derivatives. Before solving the equations of motion \eqref{Cotten_tensor}, we derive the necessary boundary conditions required to satisfy the variational principle associated with action \eqref{CSG1}.
\subsection{The variational principle}
We work with coordinates $\{r, x^+,x^-\}$ and consider the metric in Fefferman Graham gauge,
\begin{align}
	ds^2=\frac{l^2}{r^2}\,dr^2+\frac{r^2}{l^2}\,g_{ab}\,dx^adx^b, \label{FGnor}
\end{align} 
where $0\leq r < \infty$ and the boundary coordinates  $x^{\pm}= t \pm l \phi $ with $-\infty < t < +\infty$ is the time-like coordinate and $0 \leq \phi < 2\pi$. The dimensionful parameter $l$ will turn out to be related to the curvature of the locally $AdS_3$ geometries such as those of \cite{Avery:2013dja}, that are compatible with boundary conditions consistent with our variational principle to be derived later in this section.  The variation of  the action \eqref{CSG1} yields
\begin{align}
	\delta S=\frac{k}{2} \int d^3 x\, \sqrt{-G}\, C^{\mu\nu}\delta G_{\mu\nu}+ \int d^2x\, \sqrt{-\gamma}\,n_{\mu}\theta^{\mu} \label{varCSG}
\end{align}
where $n_{\mu}=-\frac{l}{r}\,\delta^r_{\mu}$ is normal to the hypersurface  $r=r_{0}$ and $\gamma_{ab}$ is the induced metric on this hypersurface.  The presymplectic potential is given by
\begin{align}
	\theta^{\mu}=\frac{k}{2} \epsilon^{\lambda\mu\nu}\,\Gamma^{\alpha}_{\lambda \rho}\,\delta\Gamma^{\rho}_{\alpha\nu}+ \frac{k}{2}\,\epsilon^{\lambda\sigma\nu}\,R^{\rho \mu}_{~~\nu\sigma}\,\delta G_{\lambda \rho} \, .\label{unmodtheta}
\end{align}
 Using the relation $\sqrt{-\gamma}=\frac{r}{l}\sqrt{-G}$, the boundary term in \eqref{varCSG} can be written as
\begin{align}
	\int d^2 x\,\left(-\delta^r_{\mu}\, \theta^{\mu}\right)&=-\int d^2 x\,\frac{k}{2} \left[\hat{\epsilon}^{\lambda r \nu}\,\Gamma^{\alpha}_{\lambda \rho}\,\delta\Gamma^{\rho}_{\alpha\nu}+ \hat{\epsilon}^{\lambda\sigma\nu}\,R^{\rho r}_{~~\nu\sigma}\,\delta G_{\lambda \rho}\right] \cr 
	&=\int d^2x\,k\,\left[\frac{\hat{\epsilon}^{ab}}{2}\,\Gamma^{\alpha}_{a\rho}\delta \Gamma^{\rho}_{\alpha b}- \hat{\epsilon}^{ab}\,R^{cr}_{~~r b}\,\delta G_{ac}\right] \label{full_CSG_bdryterm}
\end{align}
where we set $\hat{\epsilon}^{rab}=\hat{\epsilon}^{ab}$. The first term can be expanded as
\begin{align}
	\hat{\epsilon}^{ab}\,\Gamma^{\alpha}_{a\rho}\delta \Gamma^{\rho}_{\alpha b}=\hat{\epsilon}^{ab}\left[\Gamma^r_{ac}\,\delta\Gamma^c_{rb}+\Gamma^c_{ar}\,\delta\Gamma^r_{cb}+\Gamma^c_{ad}\delta\Gamma^d_{cb}\right].\label{CSG_bdry_term}
\end{align}
\begin{comment}
	The variation of the action at the boundary $r=r_0$ after imposing the equation of motion is, 
	\begin{align}
		\delta S =-\frac{k}{2} \int d^2 x\, \frac{l}{r}\epsilon^{ij}\,\left[\Gamma^r_{ik}\,\delta\Gamma^k_{rj}+\Gamma^k_{ir}\,\delta\Gamma^r_{kj}+\Gamma^k_{il}\delta\Gamma^l_{kj}\right] -k\,\int d^2 x\,\frac{l}{r}\,\epsilon^{ij}\,R^{k r}_{~~r j}\,\delta g_{ik}  \label{boundary_term}
	\end{align}
\end{comment}
%
In arriving at the above expression, we have used $\Gamma^r_{a r}=0$.
Using the metric \eqref{FGnor} the first two terms in \eqref{CSG_bdry_term} have only $r$ derivatives on the variation of the boundary metric. The third term $	\hat{\epsilon}^{ab}\,\Gamma^c_{ad}\,\delta\Gamma^d_{cb}$ can be written as
\begin{align}
	\hat{\epsilon}^{ab}\,\Gamma^c_{ad}\,\delta\Gamma^d_{cb}=&\left[\partial_e \left(\epsilon^{ab}\Gamma^e_{ad}\,g^{dc}\delta g_{cb}\right)+\partial_b \left(\epsilon^{ab}\Gamma^e_{ac}\,g^{cd}\delta g_{de}\right)-\partial_d \left(\epsilon^{ab}\Gamma^e_{ac}\,g^{cd}\delta g_{eb}\right)\right] \nonumber \\
	&+\epsilon^{ab}\,g^{de}\left[-\nabla_c\Gamma^c_{ad}\,\delta g_{eb}-\nabla_b\Gamma^c_{ad}\,\delta g_{ce}+\,\nabla_e\Gamma^c_{ad}\,\delta g_{cb}\right]. \label{bdryterm3}
\end{align}
Dropping the total derivative terms we will only be left with the second line of \eqref{bdryterm3}. Near the boundary, one  expands the metric $g_{ab}$ in \eqref{FGnor} as 
\begin{align}
	g_{ab}=g^{(0)}_{ab}+\frac{l}{r}g^{(1)}_{ab}+\frac{l^2}{r^2}\,g^{(2)}_{ab}+\cdots\, .
\end{align}
Substituting this in \eqref{full_CSG_bdryterm} and expanding in $r$, one can check that at $\mathcal{O}(r)$  the first two terms in \eqref{CSG_bdry_term} cancel the terms coming from the second term in \eqref{full_CSG_bdryterm}, and therefore, there is no divergent boundary term in the variation of the action when  \eqref{full_CSG_bdryterm} is expanded near the boundary.
At $\mathcal{O}(1)$ we find the following terms,
\begin{align}
	&\hat{\epsilon}^{cb}\left(\frac{1}{l^2}\,g^{(0)}_{dc}\,g_{(2)}^{ad}-\frac{1}{4l^2}\,g_{(0)}^{ef}g^{(1)}_{ef}\,g^{(0)}_{dc}g_{(1)}^{ad}\right)	\delta g^{(0)}_{ab}-\hat{\epsilon}^{cb}\left(\frac{1}{4l^2}\,g^{(0)}_{dc}g_{(1)}^{ad}\delta g^{(1)}_{ab}\right) \cr & -\frac{\hat{\epsilon}^{ab}}{4}\,g_{(0)}^{ed}\,\left(D_c\tilde{\Gamma}^c_{ad}\,\delta g^{(0)}_{eb}+D_b \tilde{\Gamma}^c_{ae}\,\delta g^{(0)}_{cd}-D_e \tilde{\Gamma}^{c}_{ad}\delta g^{(0)}_{cb}\right) \label{final_bdry_term_CSG}
\end{align}
where $D_a$ is the covariant derivative associated with $g^{(0)}_{ab}$ and $\tilde{\Gamma}^a_{bc}$ are Christoeffel symbols associated with it. We parameterise our boundary metric $g^{(0)}_{ab}$ in Polyakov's chiral gauge \cite{Polyakov:1987zb} such that the boundary line element is given by
\begin{align}
	ds^2_{bdry}=-\,dx^+ dx^-+g^{(0)}_{++}(x^+,x^-)(dx^+)^2 \, .   \label{bdry_met_pol}
\end{align}
With this choice of boundary metric the first term in \eqref{final_bdry_term_CSG} becomes
\begin{align}
	\frac{1}{2l^2}\,g^{++}_{(2)}\,\delta g^{(0)}_{++} +\frac{1}{8 l^2}\, g^{ef}_{(0)} g^{(1)}_{ef}\,g^{++}_{(1)}\delta g^{(0)}_{++} \,  .\label{var-first}
\end{align}
Following \cite{Compere:2013bya, Avery:2013dja}, we set $g^{++}_{(2)}=g^{(2)}_{--}=-\frac{1}{4}$ such that global AdS$_3$ geometry is part of the solution space. Therefore, the first term in \eqref{var-first} becomes $$-\frac{1}{8l^2}\,\delta^{a}_{+}\delta^{b}_{+}\,\delta g^{(0)}_{ab} $$ which can be written as a total variation $\delta \left(-\frac{1}{8l^2}\,\delta^{a}_{+}\delta^{b}_{+}\, g^{(0)}_{ab}\right) $. To cancel this term we follow the prescription of \cite{Avery:2013dja}, where one adds the following boundary term to the action \eqref{CSG1} 
\begin{align}
	S'=k\,\int d^2 x\,\mathcal T^{ab}\, g^{(0)}_{ab} ~~ \text{with} ~~
	\mathcal T^{ab}=\frac{1}{8 l^2}\,\delta^a_+\,\delta^b_+ 
\end{align}
such that from the variation of the total action $\delta S_{total}=\delta S_{CSG}+\delta S'$ the term $-\frac{1}{8l^2}\,\delta^{a}_{+}\delta^{b}_{+}\,\delta g^{(0)}_{ab} $ drops out. The trace of $g_{(1)}$ with respect to $g_{(0)}$ in the second term in \eqref{var-first} will vanish for the gauge choice \eqref{Weyl_gauge} and other conditions coming from variational principle. Now we come to the second term in  \eqref{final_bdry_term_CSG} which becomes
\begin{align}
	\frac{1}{2l^2}\,\left(g^{(1)}_{++}\,\delta g^{(1)}_{--}-g^{(1)}_{--}\,\delta g^{(1)}_{++}\right)+\frac{1}{l^2}\,g^{(0)}_{++}\left(g^{(1)}_{+-}\,\delta g^{(1)}_{--}-g^{(1)}_{--}\,\delta g^{(1)}_{+-}\right).
\end{align}
There is a clear choice of boundary conditions for which all of the above terms will go to zero and that is $g^{(1)}_{--}= \delta g^{(1)}_{--}=0$. Also due to the chiral conditions imposed on the boundary metric as in \eqref{bdry_met_pol} all the terms in the second line of \eqref{final_bdry_term_CSG} vanish. To summarise, we solve the variational problem as follows:
\begin{itemize}
	\item Parameterise the boundary metric in the Polyakov gauge as
	\begin{align}
		g^{(0)}_{+-}=-\frac{1}{2}\,,g^{(0)}_{--}=0\,,~~g^{(0)}_{++}=g^{(0)}_{++}(x^+,x^-).
	\end{align}
	\item Fix the value of $g^{(2)}_{--}=-\frac{1}{4}$ following \cite{Avery:2013dja} and add a boundary term $S'=k\int d^2 x\,\mathcal T^{ab}\, g^{(0)}_{ab}$ to the action \eqref{CSG1}.\footnote{This boundary action will transform $\theta \rightarrow \theta+ \delta M$ such that symplectic potential $\omega$ \eqref{LW} remains unchanged, therefore, it won't contribute to the final charge. }
	\item Impose the condition $g^{(1)}_{--}=0$.
\end{itemize}
\subsection{Solutions to conformal gravity} 
Now we solve the equation $C_{\mu\nu}=0$ as an expansion near the boundary at each order in $1/r$ after imposing boundary conditions derived in the previous section.   Writing the metric in Fefferman-Graham gauge as follows:
\begin{align}
	ds^2=\frac{l^2}{r^2}\,dr^2+\frac{r^2}{l^2}\,\left(g^{(0)}_{ab}+\frac{l}{r}\,g^{(1)}_{ab}+\frac{l^2}{r^2}\,g^{(2)}_{ab}+\cdots\right)dx^a\,dx^b, \label{FG1}
\end{align}
 we further impose the additional gauge condition 
\begin{align}
	g_{AdS_3}^{\mu \nu} G_{\mu \nu}= 	g_{AdS_3}^{\mu \nu}(g_{AdS_3})_{\mu \nu} =3 \label{Weyl_gauge}
\end{align}
to fix Weyl symmetry, where $(g_{\mu \nu})_{AdS_3}$ is the metric of global AdS$_3$ given by the following line element
\begin{align}
	(ds^2)_{AdS_3}=\frac{l^2}{r^2}dr^2-\frac{r^2}{l^2}\,dx^+\,dx^--\frac{1}{4}\left((dx^{-})^2+(dx^{+})^2\right)-\frac{l^2}{16\,r^2}dx^+\,dx^{-}.
\end{align}
 The condition \eqref{Weyl_gauge} sets off-diagonal components of boundary metric to be,
 \begin{align}
 &g_{+-}=\frac{1}{32 r^4 \left(l^4+16 r^4\right)}\left(128 l^2\,r^6\,g_{--}(r,x^a)+128 l^2\,r^6\,g_{++}(r,x^a)-(l^4-16 r^4)^2\right)	\cr
 & \implies g^{(0)}_{+-}=-\frac{1}{2} \end{align}
   The expansion \eqref{FG1} is similar to that of \cite{Compere:2013bya, Avery:2013dja, Troessaert:2013fma} except for the presence of  $g^{(1)}_{ab}$ and other terms that appear at odd powers of $\frac{l}{r}$. We impose chiral boundary conditions that satisfy our variational principle:
\begin{align*}
	g^{(0)}_{--}=0\,,~g^{(1)}_{--}=0\,,~g^{(2)}_{--}=-\frac{1}{4}. \label{CG_boundary}
\end{align*}
With these boundary conditions,the solution to $C_{\mu \nu}=0$ for first few orders in $r$ are given as follows :
\begin{align} g^{(2)}_{++}&=T(x^+)+\frac{1}{2}l^2\,\partial_{-}\partial_{+}g^{(0)}_{++}-\frac{1}{2}\,l^2\,(\partial_{-}g^{(0)}_{++})^2+l^2\,g^{(0)}_{++}\,\partial_{-}^2g^{(0)}_{++}\,,~~~~~g^{(3)}_{--}=0 \cr
g^{(3)}_{++}&=\frac{1}{12}\bigg(g^{(0)}_{++}\,\left(-5\,g^{(1)}_{++}+8\,l^2\,\partial_{-}^2 g^{(1)}_{++}\right)+4l^2\,\left(-\partial_{-}g^{(0)}_{++}\partial_{-}g^{(1)}_{++}+g^{(1)}_{++}\,\partial_{-}^2g^{(0)}_{++}\right) \cr&~~~~~~~~~~+4l^2\,\partial_{+}\partial_{-}g^{(1)}_{++}\bigg)\,,~~~~~g^{(4)}_{--}=-\frac{l^2}{8}\,\partial^2_{-}g^{(0)}_{++}
\end{align}
The metric components at higher order in $r$ in \eqref{FG1} are given in terms of free data $\{g^{(0)}_{++}(x^+,x^-), g^{(1)}_{++}(x^+,x^-), T(x^+)\}$, where $g^{(0)}_{++}(x^+,x^-),g^{(1)}_{++}(x^+,x^-) $ satisfy the following constraint equations. 
\begin{align}
	& \partial_{-}g^{(0)}_{++}(x^+,x^-)+l^2\,\partial^3_{-}g^{(0)}_{++}(x^+,x^-)=0 ,\nonumber \\
	&g^{(1)}_{++}(x^+,x^-)+4\,l^2\,\partial^2_{-}g^{(1)}_{++}(x^+,x^-)=0 .\label{constraint_CG}   \end{align}
These are solved as
\begin{align}
	&g^{(0)}_{++}=J_{-}(x^+)\,e^{-\frac{i\,x^-}{l}}+J_{0}(x^+)+J_{+}(x^+)\,e^{\frac{i\,x^-}{l}} = \sum_{a=0,\pm 1} J_{a}(x^+)\,\,e^{a\,\frac{i\,x^-}{l}},\,\\
	&g^{(1)}_{++}= G_{-1/2}(x^+)\,e^{-\frac{i\,x^-}{2\,l}}+G_{1/2}(x^+)\,e^{\frac{i\,x^-}{2\,l}} =\sum_{s=\pm 1/2} G_s(x^+)\,\,e^{s\,\frac{i\,x^-}{l}}. \label{g1++}
\end{align}
Thus the complete solution space for  $C_{\mu\nu}=0$  is  characterised by six chiral functions $\{J_a(x^+), G_s(x^+), T(x^+)\}\,,~~a\in \{0,\pm 1\} \text{~and}~ s \in \{\pm \frac{1}{2}\}$. 
\subsection{Asymptotic symmetries}
Since the conformal gravity \eqref{CSG1} is diffeomorphism and Weyl invariant, the asymptotic asymmetries are the combinations of residual diffeos $(\xi)$ and Weyl transformations $(\sigma)$
\begin{align}
\label{diffplusweyl}
	\delta G_{\mu\nu}=\nabla_{\mu}\xi_{\nu}+\nabla_{\nu}\xi_{\mu}-2\sigma G_{\mu\nu}
\end{align}
that leave our solution space form-invariant. Imposing the gauge condition $\delta G_{rr}=0$ leads to the determination of Weyl parameter $\sigma$ as \footnote{Note that $\sigma$ corresponds to bulk Weyl rescaling and is an additional gauge symmetry of the solutions of conformal gravity. This is in contrast with Einstein gravity  where $\sigma$ is the boundary Weyl rescaling parameter.  }, 
%on the radial component of the vector field $\xi$ 

% 
\begin{align}
	\sigma=-\frac{\xi^r}{r}+\partial_r \xi^r.
\end{align}
We further expand the vector field component near the boundary as,
\begin{align}
	&\xi^r=r^2\,\xi^r_{(-2)}(x^a)+r\,\xi^r_{(-1)}(x^a)+\xi^r_{(0)}(x^a)+\cdots \, ,\cr 
	&\xi^a= \sum_{n=0}^\infty \,r^{-n}\xi^a_{(n)}(x^b) \, .
\end{align}
From the gauge condition $\delta G_{ra}=0$ at $\mathcal{O}(1)$, one gets
\begin{align}
	\xi^a_{(1)}=l^4\,g_{(0)}^{ab}\,\partial_b \xi^r_{(-2)} .  
\end{align}
From $\delta G_{--}=0$ at $\mathcal{O}(r^2)$ and $\mathcal{O}(r)$, one gets
\begin{align}
	\xi^{+}_{(0)}=\lambda(x^+)\,,~~\frac{1}{2}\,\xi^r_{(-2)}+2\,l^2\,\partial^2_{-}\xi^r_{(-2)}=0  .\label{vec_cons1}   
\end{align}
The equation \eqref{vec_cons1} can be solved for $\xi^r_{(-2)}$ to find,
\begin{align}
	\xi^r_{(-2)}=\chi_{1/2}(x^+)\,\,e^{\,\frac{i\,x^-}{2l}}+\chi_{-1/2}(x^+)\,\,e^{\,-\frac{i\,x^-}{2l}}.
\end{align}
From $\delta G_{+-}=0$ at $\mathcal{O}(r^2)$ one obtains
\begin{align}
	\xi^{r}_{(-1)}=-\frac{1}{2}D_a \xi^a_{(0)}  .
\end{align}
The gauge condition $\delta G_{r-}=0$ at the $\mathcal{O}\left(r^{-1}\right)$ gives
\begin{align}
	\xi^{+}_{(2)}=-l^4\,\partial_{-}\xi^r_{(-1)}.
\end{align}
Using this at $\delta G_{--}=0$ at $\mathcal{O}(r^0)$ one obtains the following constraint equation,
\begin{align}
	\partial_{-}\xi^{-}_{(0)}+l^2\,\partial^3_{-}\xi^{-}_{(0)}=0   
\end{align}
which can be solved as
\begin{align}
	\xi^{-}_{(0)}=\lambda_{-}(x^+)\,e^{-\frac{i\,x^-}{l}}+\lambda_{0}(x^+)+\lambda_{+}(x^+)\,e^{\frac{i\,x^-}{l}}.
\end{align}
Therefore, the final residual diffeomorphisms and Weyl transformations correspond to
\begin{align*}
	&\xi^{+}=\lambda(x^+)+\cdots\,,~~\xi^{-}= \sum_{a=\pm 1,0} \lambda_{a}(x^+)\,\,e^{a\,\frac{i\,x^-}{l}}+\cdots~~~ ,\nonumber\\
	&\xi^r=r^2\,\sum_{s=\pm 1/2} \chi_s(x^+)\,\,e^{s\,\frac{i\,x^-}{l}}+\cdots\,~~.
\end{align*}
These induce the following variations on the background fields $(T(x^+), J^a(x^+), G_s(x^+))$ in the solutions
\begin{align}
	&\delta J_{a}=\lambda\,\partial J_a+\partial \lambda J_a+\partial \lambda_a+i\,{f_{a}}^{bc}\,J_{b}\lambda_c-4\,l^3\,\left(\lambda_a\right)^{ss'}\,G_s\,\chi_{s'},\cr &\cr		
	&\delta G_{s}=\lambda \partial G_s+\frac{3}{2}G_s\,\partial\lambda-\partial^2\chi_s-(T+\frac{1}{2}\eta^{ab}\,J_aJ_b)\chi_s  \cr &~~~~~~~~~~~~~~~~~~~~+i\left(\lambda^a\right)^{s'}_{~s}\left(2\,J_a\,\partial \chi_{s'}+\partial J_a\,\chi_{s'}+\lambda_a\,G_{s'}\right),\cr&\cr
	&\delta T=2\,T\,\partial \lambda+\lambda\,\partial T+\frac{1}{2}\,\partial^3 \lambda-\frac{1}{2}\eta^{ab}J_a\,\partial \lambda_b-i\,l^3\,\epsilon^{ss'}\,\left(3\,G_s\,\partial \chi_{s'}+\partial G_s\,\chi_{s'}\right) .\label{trans_CSG}\cr
\end{align}
To obtain these variations we do the following redefinition of background fields,
\begin{align}
	&J_+\rightarrow -\frac{l}{2}J_+,~~J_-\rightarrow -\frac{l}{2} J_{-},~~ J_0\rightarrow -l\,J_0\,,~~ C_s \rightarrow -2 l^3\,C_s\,,~~T \rightarrow -l^2\,T \cr
	&\lambda_{-}\,\rightarrow \frac{l}{2}\,\lambda_{-}\,,~~\lambda_0\rightarrow l\,\lambda_0\,,~~\lambda_+ \rightarrow \frac{l}{2}\lambda_+.
\end{align} 
\subsection{Calculation of Charges}
We now turn to calculating the charges that induce variations of the background field \eqref{trans_CSG} due to residual diffeomorphisms and Weyl symmetries. As mentioned before, the Lagrangian density is invariant up to the total derivative under (\ref{diffplusweyl}) where $\xi^\mu$ are the diffeomorphism parameters and $\sigma$ is the Weyl transformation one.
To calculate the charges we use modified covariant phase space formalism proposed by Tachikawa \cite{Tachikawa:2006sz} which we review in Appendix \ref{rev_CSG}.\footnote{See also \cite{Dengiz:2020fpe} for a derivation of charges associated with Weyl-invariant Topologically Massive Gravity.} 
%The symmetry parameters for our theory are $\chi \in \{\xi, \sigma\}$.
Let us calculate the charges associated with the residual Weyl symmetry first. Following \eqref{mod_lag}, the variation of Lagrangian density under Weyl transformation $(\sigma)$ is given by
\begin{align}
	\delta_{\sigma}L= \partial_{\mu}\Xi^{\mu}_{(\sigma)}
\end{align}
where 
\begin{align}
	&\Xi^{\mu}_{(\sigma)}=\frac{k}{2}\,\hat{\epsilon}^{\lambda\mu\nu}G^{\alpha\beta}\,\partial_{\alpha}\sigma\,\partial_{\lambda}G_{\beta\nu}. \label{weyl_var}  
\end{align}
 Under Weyl transformation, the symplectic potential \eqref{unmodtheta} transforms as follows,
\begin{align}
	\delta_{\sigma}\theta^{\mu}&=\Pi^{\mu} \cr
	\Pi^{\mu}&=\frac{k}{2}\hat{\epsilon}^{\lambda\mu\nu}\big(-\partial_{\lambda}\sigma\delta\Gamma^{\alpha}_{
		\alpha\nu}+\partial^{\alpha}\sigma\,\partial_{\lambda}\delta G_{\alpha\nu}+\partial_{\beta}\sigma\,\partial_{\lambda}G_{\nu\rho}\,\delta G^{\beta\rho}\big) \nonumber\\ &+\frac{k}{2}\,\epsilon^{\lambda \mu \nu}\,\partial_{\nu}\left(2\,\delta \Gamma^{\alpha}_{\alpha \lambda}\,\sigma+2\,\delta G_{\alpha \lambda}\,\partial^{\alpha}\sigma\right)
\end{align}
Here we have used the identity $\frac{k}{2}\,\hat{\epsilon}^{\lambda \sigma \nu} R^{\rho \mu}_{~~\nu \sigma} \, \delta G_{\rho\lambda}=-k\,\hat{\epsilon}^{\lambda \mu\nu}\,G^{\alpha \beta }R_{\beta \lambda}\,\delta G_{\nu \alpha}$. Now from the formula \eqref{totaldiv} one obtains
\begin{align}
	\Sigma^{\mu \nu}_{(\sigma)}=-k\,\epsilon^{\lambda \mu \nu}\,\delta G_{\alpha \lambda}\,\partial^{\alpha}\sigma-\frac{k}{2}\,\epsilon^{\lambda \mu \nu}\,\delta \Gamma^{\alpha}_{\alpha \lambda}\,\sigma \label{other_termsinQ}
\end{align}
The second term in the \eqref{other_termsinQ} vanishes on-shell. The Weyl charge  using the formula \eqref{final_two_form} is given by,
\begin{align}Q^{\mu \nu}_{(\sigma)}= \frac{1}{16 \pi}\left(-k\,\hat{\epsilon}^{\mu \nu\lambda} \partial_{\alpha}\sigma\,G^{\alpha \beta} \delta G_{\lambda \beta}\right).
\end{align}
 Under diffeomorphisms $(\xi)$, the Lagrangian density transform as follows,
 \begin{align}
 	\delta_{\xi}L= \partial_{\mu}\left(\xi^{\mu}\,L\right)+\partial_{\mu}\Xi^{\mu}_{(\xi)}
 \end{align} 
 where
 \begin{align}
 	\Xi^{\mu}_{(\xi)}= \frac{k}{2}\,\hat{\epsilon}^{\lambda \mu \nu}\,\partial_{\nu}\Gamma^{\beta}_{\lambda \alpha}\,\partial_{\beta}\xi^{\alpha} \label{diff_var}
 \end{align}
Due to the presence of second term in \eqref{diff_var}, the Lagrangian is not covariant under the diffeomorphisms. The calculation of the charges for diffeomorphism symmetries has been carried out in \cite{Tachikawa:2006sz, Dengiz:2020fpe, Kim:2013cor} in the context of topologically massive gravity. For diffeomorphisms following \cite{Dengiz:2020fpe} and using \eqref{diff_var}, one finds
\begin{align}
 \Sigma^{\mu\nu}_{(\xi)}=\frac{k}{2}\,\hat{\epsilon}^{\lambda \mu \nu}\,\delta \Gamma^{\beta}_{\lambda \alpha}\partial_{\beta}\xi^{\alpha},
\end{align}
and the final diffeomorphism charge is 
\begin{align}
	16 \pi Q^{\mu \nu}_{\xi}=-2\,k\,\hat{\epsilon}^{\mu \nu\rho}\left(\delta S_{\rho \alpha}\xi^{\alpha}-\frac{1}{2}\delta \Gamma^{\alpha}_{\rho \beta}\,\nabla_{\alpha}\xi^{\beta}+\xi^{\beta}\,R^{\alpha}_{[\beta}\delta G_{\rho]\alpha}\right).
\end{align}
Therefore, the total charge becomes
\begin{align}
	16 \pi (Q^{\mu \nu}_{\xi}+ Q^{\mu \nu}_{(\sigma)})= -2\,k\,\hat{\epsilon}^{\mu \nu\rho}\left(\delta S_{\rho \alpha}\xi^{\alpha}-\frac{1}{2}\delta \Gamma^{\alpha}_{\rho \beta}\,\nabla_{\alpha}\xi^{\beta}+\xi^{\beta}\,R^{\alpha}_{[\beta}\delta G_{\rho]\alpha}+\frac{1}{2}\partial_{\alpha}\sigma\,G^{\alpha \beta} \delta G_{\rho \beta}\right).
\end{align}
We integrate this charge on a constant $t$ surface at the boundary and use the coordinates $x^\pm  = t \pm  l \phi$ to obtain,
\begin{align}
	16 \pi \cancel{\delta}Q=\int_{0}^{2\pi} d\phi\,\, Q^{r t} =\int_{0}^{2\pi} d\phi \,\frac{1}{2}\left(Q^{r+}+Q^{r-}\right).
\end{align}
The Weyl charges $Q^{\mu\nu}_{(\sigma)}$ cancel the divergent piece at $\mathcal{O}(r)$ from the diffeomorphism charges $\mathcal{Q}^{\mu\nu}_{\xi}$ and the resultant charge is given by
\begin{align}
	16 \pi \cancel{\delta}Q=\frac{k}{2}\int_0^{2\pi} d\phi \left(2\,\delta T(x^+)\,\lambda-\,\eta^{ab}\delta J_{a}\,\lambda_b-4\,i\,l^3\,\epsilon^{ss'}\,\delta G_s\,\chi_{s'}\right)+  \int_{0}^{2\pi}d\phi\,\partial_{\phi}A(x^+,x^-) . \label{varcharge}
\end{align}
where,
\begin{align}
	A(x^+,x^-)=-\frac{k\,l}{2}\,\partial_{-}\left(e^{-\frac{ix^-}{l}}\,\delta J_{-1}(x^+)\,\lambda(x^+)+e^{\frac{i x^-}{l}}\,\delta J_{1}(x^+)\,\lambda(x^+)\right)
\end{align}
This charge is integrable and the second term in \eqref{varcharge} drops out as it is a total derivative. Therefore the final charge is
\begin{align}
	Q=\frac{k}{32\pi}\int_0^{2\pi} d\phi\left(2\,T(x^+)\,\lambda(x^+)-\eta^{ab}\,J_a\,\lambda_b-4\,i\,l^3\,\epsilon^{ss'}\, G_s\,\chi_{s'}\right) \, . \label{secondordercharge}
\end{align}
Using the fact that this charge $Q_\Lambda$ is expected to induce the change $\delta_\Lambda f = \{ Q_\Lambda, f\}$, one can read out brackets among $(T(\phi), J_a(\phi), G_r(\phi))$. This results in
\begin{align}
	\frac{k}{16\,\pi} \{ T(\phi), T(\psi) \} &= 2 \, T(\psi)  \, \delta'(\psi-\phi) + \delta(\psi-\phi) \, T'(\psi)   + \frac{1}{2} \delta^{'''}(\psi -\phi) ,\cr 
	& \cr
	\frac{k}{16\,\pi} \{ T(\phi), J_a(\psi) \} &=  J_a(\psi) \, \delta'(\psi-\phi) +  J_a'(\psi) \, \delta(\psi - \phi) ,\cr
		& \cr
	\frac{k}{16\pi} \{ T(\phi), G_r(\psi) \} &= \frac{3}{2} G_r(\psi) \, \delta' (\psi - \phi) +  G_r'(\psi) \, \delta(\psi -\phi) , \cr
	 \label{Poissbrac1}
\end{align}
\begin{align}
	& \cr
	-\frac{k}{32\pi}\{ J_a(\phi), J_b(\psi) \} &= i {f_{ab}}^c J_c(\psi) \, \delta (\psi-\phi) + \eta_{ab} \, \delta'(\psi-\phi) ,\cr
	& \cr
	-\frac{k}{32\pi}\{ J_a(\phi), G_r(\psi) \} &= i {(\lambda_a)^s}_r G_s(\psi) \, \delta(\psi -\phi) , \cr
	& \cr
	\frac{k l^3}{16\pi} \{ G_r(\phi), G_s(\psi) \} &= \frac{i}{2} \epsilon_{rs} \, \left(T (\psi)  + \frac{1}{2} J^2(\psi) \right) \, \delta(\psi-\phi) + \frac{i}{2} \epsilon_{rs} \, \delta''(\psi-\phi) \cr
	& + (\lambda^a)_{rs} \left[ J_a(\psi) \, \delta'(\psi-\phi) + \frac{1}{2} J_a'(\psi) \, \delta(\psi-\phi) \right] \label{Poissbrac2}
\end{align}
where all derivatives denoted by primes are with respect to the arguments. We use the mode expansions 
\begin{align} 
	T(\phi) = \frac{1}{2\pi} \sum_{n \in {\mathbb Z}} T_n \, e^{i n \phi} &\iff L_n =  \int_0^{2\pi} d\phi \, T(\phi) e^{-in\phi}, \cr
	J_a(\phi) = \frac{1}{2\pi} \sum_{n \in {\mathbb Z}} J_{a,n} \, e^{i n \phi} &\iff J_{a,n} =  \int_0^{2\pi} d\phi \, J_a(\phi) e^{-in\phi}, \cr
	G_s(\phi) = \frac{1}{2\pi} \sum_{r \in {\mathbb Z}+\frac{1}{2} } G_{s,r} \, e^{i r \phi} &\iff G_{s,r} =  \int_0^{2\pi} d\phi \, G_s(\phi) e^{-ir\phi} \, .
\end{align}
Note that we have considered anti-periodic boundary conditions for $G_r(\phi)$. This condition was necessary to ensure that $g^{(1)}_{++}$ in \eqref{g1++} remains periodic as $\phi \rightarrow \phi+ 2\pi$. Using the mode expansion of $J_a(\phi)$ we can write the mode expansion of $\eta^{ab} J_a(\phi) J_b(\phi)$ as follows:
\begin{align}
	\eta^{ab} J_a(\phi) J_b(\phi) &= \frac{1}{4\pi^2}\sum_{m,n \in {\mathbb Z}} \eta^{ab} J_{a,m} J_{b,n} \, e^{i (m+n)\phi} \cr
	&= \frac{1}{4\pi^2}\sum_{m+n=p \in {\mathbb Z}} \left( \sum_{m \in {\mathbb Z}} \eta^{ab} J_{a,m} J_{b,p-m} \right) \, e^{i p\phi} :  = \frac{1}{4\pi^2} \sum_{p \in {\mathbb Z}}  (J^2)_p  ~e^{i p\phi} \cr
	&\implies  (J^2)_p = 2\pi \int_0^{2\pi} d\phi \, \eta^{ab} J_a(\phi) J_b(\phi) \, e^{-ip\phi} . 
\end{align}
\begin{comment}
	\begin{align}
		&\frac{k}{16\pi}\{T(\phi),T(\psi)\}=2\,T(\psi)\,\delta^{'}(\psi-\phi)+\delta(\psi-\phi)\,T^{'}(\psi)+\frac{1}{2}\delta^{'''}(\psi-\phi)\cr
		&\frac{k}{16\pi}\{T(\phi),J_a(\psi)\}=J_a(\psi)\,\delta^{'}(\psi-\phi)+J_{a}(\psi)^{'}\delta(\psi-\phi)\cr
		&\frac{k}{16\pi}\{T(\phi),C_s(\psi)\}=\frac{3}{2}C_s(\psi)\,\delta^{'}(\psi-\phi)+C_s(\psi)'\delta(\psi-\phi)\cr
		&-\frac{k}{32\pi}\{J_a(\phi),J_b(\psi)\}=i f_{ab}^{~~c}\,J_c(\psi)\,\delta(\psi-\phi)+\eta_{ab}\,\delta^{'}(\psi-\phi)\cr
		&-\frac{k}{32\pi}\{J_a(\phi),C_s(\psi)\}=i\left(\lambda_a\right)^{s'}_{~~s}\,C^s(\psi)\,\delta(\psi-\phi)\cr \cr
		&\frac{k\,l^3\,N}{8\pi}\{C_{s'}(\phi),C_{s}(\psi)\}=i\,\epsilon_{s's}\left(T(\psi)+\frac{1}{2}J^2(\psi)\right)\delta(\psi-\phi)+i\,\epsilon_{s's}\,\delta^{''}(\psi-\phi)\nonumber \\
		&\left(\lambda^a\right)_{s' s}\left(2\,J_{a}\delta^{'}\,(\psi-\phi)+(J_{a})^{'}\delta(\psi-\phi)\right)
	\end{align}
\end{comment}
Replacing the Poisson brackets $\{A, B\}$ in terms of the Dirac brackets $[A, B]$ using $\{A, B\} = -i [A, B]$ and $L_n \rightarrow \frac{16\pi}{k}\, L_n$, $J_{a,n} \rightarrow \frac{32\pi}{k} J_{a,n}$ and $G_{s,r} \rightarrow \frac{4\pi}{\sqrt{l^3 k}} G_{s,r}$ the resultant commutation relations are 
\begin{align}
	&\left[L_m,L_{n}\right]=(n-m)L_{m+n}+\frac{k}{8} n^3\,\delta_{m+n,0} \, ,\\
	& \left[L_{m},J_{a,n}\right]=-n\,J_{a,n+m}, ~~\left[L_{m},C_{s,r}\right]=\frac{1}{2}(m-2r)C_{s,m+r} \, ,\\
	&\left[J_{a,m},J_{b,n}\right]=f_{ab}^{~c}J_c+\frac{k}{8}\,n\,\eta_{ab}\delta_{m+n,0}\,,~~\left[J_{a,m}\,, G_{s,r}\right]=\frac{1}{2}(a-2s)\,G_{s+a,r+m} \, ,\cr 
	&\left[G_{s,r'},G_{s',r'}\right]=\epsilon_{ss'}\left[\left(r^2-\frac{1}{4}\right)\,\delta_{r+r'}-\frac{4}{k}L_{r+r'}-\frac{32}{ k^2}\left(J^2\right)_{r+r'}\right] \nonumber \\&+\frac{8}{k}\left(\lambda^a\right)_{ss'}(r-r')J_{a,r+r'} \, .\label{semi_class_GS3}
\end{align}
For $k=4\kappa$, this symmetry algebra matches with the semi-classical, large-$\kappa$, limit of the chiral $\Lambda$-$\mathfrak{bms}_4$  in \eqref{RESULT}. This proves that the gravitational Chern-Simons theory admits chiral $\Lambda$-$\mathfrak{bms}_4$ symmetry. 
\section{\texorpdfstring{Chiral $\Lambda$-$\mathfrak{bms}_4$}{Lg} from \texorpdfstring{$\mathfrak{so}(2,3)$}{Lg} CS gauge theory }
\label{first-order}

In this section, we show that the semi-classical limit of the chiral $\Lambda$-$\mathfrak{bms}_4$ is the asymptotic symmetry algebra of the 3d conformal gravity -- formulated as the Chern-Simons gauge theory for the gauge algebra $\mathfrak{so}(2,3)$ -- with a consistent set of boundary conditions.

We start with the Chern-Simons theory in 3d with gauge algebra $\mathfrak{so}(2,3)$ written as \cite{Gupta:2023fmp},
\begin{align}
	[L_m, L_n] &= (m-n) L_{m+n}, ~~ [\bar L_m, \bar L_n] = (m-n) \, \bar L_{m+n}, ~~ [L_m, \bar L_n] =0 \cr
	[L_n, G_{s,r}] &=\frac{1}{2} (n-2r) \, G_{s,n+r}, ~~ [\bar L_n, G_{s,r} ] = \frac{1}{2} (n-2s) \, G_{n+s,r}, \cr
	& ~~  [G_{s,r}, G_{s',r'}] = 2 \, \epsilon_{rr'} \, \bar L_{s+s'} + 2 \, \epsilon_{ss'} \, L_{r+r'}
\end{align}
for $m,n \cdots = 0, \pm 1$, $r,s, \cdots = \pm 1/2$. Here one may think of $L_n$ and $\bar L_n$ to generate either the $\mathfrak{so}(2,2) = \mathfrak{sl}(2, {\mathbb R}) \oplus  \mathfrak{sl}(2, {\mathbb R})$ subalgebra or $\mathfrak{so}(1,3) = \mathfrak{sl}(2, {\mathbb C}) \oplus \mathfrak{sl}(2, {\mathbb C})^c$ subalgebra of $\mathfrak{so}(2,3)$.

 The 3d gauge connection $\mathcal A$ can be written in terms of 2d gauge connection $\mathfrak a$ using gauge transformation $\mathfrak b$ as follows,
  ${\cal A} = \mathfrak{b} \, d\mathfrak{b}^{-1} + \mathfrak{b} \mathfrak{a} \mathfrak{b}^{-1} $ where $\mathfrak{b} = e^{(J_0 - L_0) \, \ln (r/{\ell})}$ and 2d coordinates are denoted by $ x^{i} \in \{x^+,x^-\}$.
We further write the gauge connection $\mathfrak{a}$ as
\begin{align}
	\mathfrak{a} = A^{(m)} L_m + {\bar A}^{(a)} J_a + \Phi^{(s,r)} G_{s,r}, \label{2dconn}
\end{align}
where $A^{(m)}, \bar A^{(a)}, \Phi^{(s,r)}$ are functions of $x^i$. Then the $3d$ flatness conditions ${\cal F} = d{\cal A} + {\cal A} \wedge {\cal A} = 0$ $\implies$ $d\mathfrak{a} +\mathfrak{a} \wedge \mathfrak{a} = 0$, in terms of the components are written as follows
\begin{align}
	\partial_i A^{(p)}_j - \partial_j A^{(p)}_i + (m-n) \delta^p_{m+n} A^{(m)}_i \, A^{(n)}_j+ 2 \epsilon_{ss'} \Phi^{(s,r)}_i \Phi^{(s',r')}_j \delta^p_{r+r'} &= 0, \cr
	\partial_i \bar A^{(c)}_j - \partial_j \bar A^{(c)}_i + (a-b) \delta^c_{a+b} \bar A^{(a)}_i \, \bar A^{(b)}_j+ 2 \epsilon_{rr'} \Phi^{(s,r)}_i \Phi^{(s',r')}_j \delta^c_{s+s'} &= 0 , \cr
	\partial_i \Phi^{(s,r)}_j  + \frac{1}{2} (m-2r') \delta^r_{m+r'} A^{(m)}_i \Phi^{(s,r')}_j + \frac{1}{2} (a-2s') \delta^s_{a+s'} \bar A^{(a)}_i \Phi^{(s',r)}_j - (i \leftrightarrow j) &= 0 .
\end{align}
 Next, we derive constraints and boundary conditions on the components of gauge connection using variational principle. 
\subsection{The constraints and a variational principle}
The action for CS gauge theory is 
\begin{align}
	S = \frac{k}{4\pi} \int_M {\rm Tr} \Big( {\cal A} \wedge d{\cal A} + \frac{2}{3} {\cal A} \wedge{\cal A} \wedge{\cal A} \Big),
\end{align}
whose variation is
\begin{align}
	\delta S 
	%&=& \frac{k}{4\pi} \int {\rm Tr} \Big( 2 {\cal F} \wedge \delta {\cal A} - d({\cal A} \wedge \delta {\cal A}) \Big) \cr
	&= \frac{k}{4\pi} \int {\rm Tr} \Big( 2 {\cal F} \wedge \delta {\cal A} \Big) - \frac{k}{4\pi} \int_{\partial M} {\rm Tr} \Big({\cal A} \wedge \delta {\cal A} \Big).
\end{align}
Around an on-shell solution ({\it i.e.,} with ${\cal F} =0$) configuration under general variation $\delta {\cal A}$ is 
\begin{align}
	\delta S \simeq - \frac{k}{4\pi} \int_{\partial M} {\rm Tr} \Big( {\cal A} \wedge \delta {\cal A} \Big). \label{boundary_term_CS}
\end{align}
Using the traces ${\rm Tr} (L_m L_n) = - \eta_{mn}$, ${\rm Tr}(J_a J_b) = -\eta_{ab}$ and ${\rm Tr}(G_{(s,r)} G_{(s',r')}) = 4 \epsilon_{ss'} \epsilon_{rr'}$  \eqref{boundary_term_CS} becomes
\begin{align}
	\delta S  \simeq &  \frac{k}{8\pi} \int_{r=r_0} \!\!\!\!\! d^2x \Big[ \eta_{mn} \left(A^{(m)}_- \delta A^{(n)}_+ -A^{(m)}_+ \delta A^{(n)}_- \right) \cr
	&+ \eta_{ab} \left(\bar A^{(a)}_- \delta \bar A^{(b)}_+ -\bar A^{(a)}_+ \delta \bar A^{(b)}_- \right) \cr
	&- 4 \, \epsilon_{ss'} \epsilon_{rr'} \left(\Phi^{(s,r)}_-\delta\Phi^{(s',r')}_+ - \Phi^{(s,r)}_+ \delta\Phi^{(s',r')}_- \right) \Big] + {\cal O}(1/r_0) \cr .
\end{align}
%
%The last term in the square-brackets vanishes if we impose, for instance: $\Phi^{(s,r)}_-  = 0$ $\forall s,r = \pm 1/2$ along with its variation. 
%
We first impose the following constraints:\footnote{The $\{+,-\}$ subscript of one-forms $\{A^{(m)},\bar{A}^{(a)}, \Phi^{(r,s)}\}$ denote the $\{x^+,x^-\}$ component respectively. In the context of field $\Phi^{(r,s)}$ even though $r,s \in \{-1/2,1/2\}$, we will denote it by $\{-,+\}$ for notational simplicity.}
\begin{align}
	\label{constraints-01}
	& A^{(-)}_- = 1, ~~ A^{(0)}_- =0,~~\bar A^{(+)}_+ = 1, ~~  \bar A^{(0)}_- = 0,\cr \
	&\Phi^{(+,-)}_+ =  \Phi^{(+,-)}_- =  \Phi^{(-,-)}_- =  \Phi^{(-,+)}_- = 0 \, ,\cr
	& \bar A^{(0)}_+ = A^{(0)}_+ \, .
\end{align}
The constraints imposed on $A^{(m)}$ and $\bar{A}^{(m)}$ in (\ref{constraints-01}) are the same as those in \cite{Poojary:2014ifa} for AdS$_3$ gravity. 
The equations of motion (flatness of ${\cal A}$) lead to the additional constraints:
\begin{align}
	\label{constraints-021}
	& A^{(0)}_+ = \partial_- A^{(-)}_+, ~~ \Phi^{(+,+)}_+ = \Phi^{(+,+)}_- A^{(-)}_+, ~~ \Phi^{(-,+)}_+ = \partial_- \Phi^{(-,-)}_+, \cr
	& \bar A^{(-)}_- = \frac{1}{2} \partial_- \bar A^{(0)}_+ - \Phi^{(-,-)}_+ \Phi^{(+,+)}_-, \cr
	& A^{(+)}_+ = \frac{1}{2} \partial_-^2 A^{(-)}_+ + A^{(-)}_+ A^{(+)}_- - \Phi^{(-,-)}_+ \Phi^{(+,+)}_-
\end{align}
and the following equations among the remaining six fields $(A^{(-)}_+, A^{(+)}_-, \bar A^{(-)}_+, \bar A^{(+)}_-, \Phi^{(-,-)}_+, \Phi^{(+,+)}_-)$ :
\begin{align}
	\frac{1}{2} \partial_-^3 A^{(-)}_+ + \left( A^{(-)}_+ \partial_-  + 2  \partial_- A^{(-)}_+ - \partial_+ \right) A^{(+)}_- &= 3 \Phi^{(+,+)}_- \partial_- \Phi^{(-,-)}_+ + \Phi^{(-,-)}_+ \partial_-\Phi^{(+,+)}_- , 
\end{align}
\begin{align}
	& \hskip 2cm \partial_- \left[\bar A^{(-)}_+ -\frac{1}{4} \left(\partial_- A^{(-)}_+ \right)^2 - \frac{1}{2} \partial_-\partial_+ A^{(-)}_+ \right] \cr
	&  + \partial_+ \left[\Phi^{(-,-)}_+ \Phi^{(+,+)}_- +\bar A^{(+)}_- \bar A^{(-)}_+ \right] + \left[\Phi^{(-,-)}_+ \Phi^{(+,+)}_- +\bar A^{(+)}_- \bar A^{(-)}_+ \right]\, \partial_-  A^{(-)}_+ =0, \cr 
\end{align}
\begin{align}
	& \partial_+ \Phi^{(+,+)}_- - A^{(-)}_+ \partial_- \Phi^{(+,+)}_- - 2 \Phi^{(+,+)}_- \partial_-A^{(-)}_+ - 2 \bar A^{(+)}_- \partial_- \Phi^{(-,-)}_+ -  \Phi^{(-,-)}_+ \partial_- \bar A^{(+)}_-=0 ,
\end{align}
\begin{align}
	& \partial_-^2 \Phi^{(-,-)}_+ - \left[\Phi^{(+,+)}_- A^{(-)}_+ + \Phi^{(-,-)}_+ \bar A^{(+)}_- \right] \left[ -\Phi^{(-,-)}_+ \Phi^{(+,+)}_- +\bar A^{(+)}_- \bar A^{(-)}_+ + \frac{1}{2} \partial_-^2A^{(-)}_+ \right]  \cr
	& + \left[ \Phi^{(+,+)}_-  \bar A^{(-)}_+ + A^{(+)}_+  \Phi^{(-,-)}_+ \right] = 0 .
\end{align}
Imposing the constraints (\ref{constraints-01}, \ref{constraints-021}), the variation around on shell solution \eqref{boundary_term_CS} becomes:
{\small 
	\begin{align}
		\delta S \simeq - \frac{k}{4\pi} \int_{r=r_0} \!\!\!\!\!\! 4 \, d\bar z \wedge dz \!\! \left[ \bar A^{(-)}_+ \delta \bar A^{(+)}_- \!-\! A^{(+)}_- \delta A^{(-)}_+  \!-\! \Phi^{(-,-)}_+ \delta \Phi^{(+,+)}_- \!+\! \Phi^{(+,+)}_- \delta  \Phi^{(-,-)}_+ \right] .
	\end{align}
}
We solve the variational problem  $\delta S \simeq 0$ in a chiral way by choosing\footnote{The first of these requires adding a further boundary term which can be done easily as in \cite{Poojary:2014ifa}.} 
\begin{align}
	\label{constraints-02}
	A^{(+)}_- = 1/4, ~~ \bar A^{(+)}_- = 0, ~~ \Phi^{(+,+)}_- = 0
\end{align}
which leads to solutions with 
\begin{align}
	A^{(-)}_+ (x^\pm ) &= J^{(-1)} (x^+) \, e^{-i x^-} + J^{(0)}(x^+) +  J^{(1)} (x^+) e^{i x^-}\cr
	\Phi^{(-,-)}_+ &= G^{(-1/2)}(x^+) \, e^{-\frac{i}{2} x^-} + G^{(1/2)} (x^+) \, e^{\frac{i}{2} x^-} \cr
	\bar A^{(-1)}_+  &= T(x^+) + \frac{1}{2} \partial_+ \partial_- A^{(-1)}_+ + \frac{1}{4} \left(\partial_- A^{(-1)}_+ \right)^2 . \label{sol_space}
\end{align}
Next we derive residual gauge transformations that preserve the constraints  (\ref{constraints-01}, \ref{constraints-02}) and leave the connection $\mathcal{A}$ form-invariant. For that we write the gauge parameter as follows,
\begin{align}
	\Lambda=\Lambda^{(m)}L_m+\bar{\Lambda}^{(a)}J_a+ \Lambda^{(s,r)}G_{s,r}, \label{gaugeparameter}
\end{align}
where $\{\Lambda^{(m)}, \bar{\Lambda}^{(a)}, \Lambda^{(s,r)}\}$ are general function of $x^{\pm}$.
The gauge transformation of the connection is given by,
\begin{align}
	\delta \mathcal A_i= \partial_i \Lambda +\left[\mathcal A_i,\Lambda\right]. \label{gaugetrans1}
\end{align}
 The solution of $\Lambda$ satisfying \eqref{gaugetrans1}  can be written in terms of functions $\{\Lambda^{(-)}, \bar{\Lambda}^{(+)}, \Lambda^{(+,-)}\}$ as follows,
\begin{align}
	&	\Lambda^{(0)}=-4\,\partial_{-}\Lambda^{(-)},~~\Lambda^{(1)}=\frac{1}{2}\lambda^{(0)}(x^+)-\frac{1}{4}\Lambda^{(-)}(x^+,x^-)\,,~\Lambda^{(+,+)}=\partial_{-}\Lambda^{(+,-)}, \nonumber\\
	&\Lambda^{(-,+)}=\frac{1}{2}\Lambda^{(+,-)}\left[J^{(-1)}(x^+)\,e^{-i x^-}+J^{(1)}(x^+)\,e^{i x^-}\right]+\partial_{-}\Lambda^{(-,-)}, \nonumber \\
	&\Lambda^{(-,-)}=\bar{\lambda}^{(1)}\,\Phi_+^{(-,-)}-\partial_+\Lambda^{(+,-)}+ \partial_-\Lambda^{(+,-)}A^{(-)}_+, \nonumber \\
	&\partial_-\bar{\Lambda}^{(0)}=-\bar{\Lambda}^{(+)}\left[J^{(-1)}(x^+)e^{-ix^-}+J^{(+1)}(x^+)e^{i x^-}\right], \nonumber \\
	&\partial_-\bar{\Lambda}^{(-)}=-\frac{1}{2}\bar{\Lambda}^{(0)}\left[J^{(-1)}(x^+)e^{-ix^-}+J^{(+1)}(x^+)e^{i x^-}\right],
\end{align} 
which are further constrained to satisfy
\begin{align}
	&\partial_{-}^2 \Lambda^{(-)}(x^+,x^-)+ \Lambda^{(-)}(x^+,x^-)-\lambda^{(0)}(x^+)=0, \nonumber \\
	& \partial_- \bar{\Lambda}^{(+)}=0\,,~~\partial_-^2\,\Lambda^{(+,-)}+\frac{1}{4} \Lambda^{(+,-)}=0. \label{constraints3}
\end{align}
These constraints can be solved to obtain
\begin{align}
	&	\Lambda^{(-)}= \lambda^{(-1)} (x^+) \, e^{-i x^-} + \lambda^{(0)}(x^+) +  \lambda^{(1)} (x^+) e^{i x^-} ,\nonumber \\
	&\bar{\Lambda}^{(+)}= \lambda(x^+) , \nonumber \\
	&\Lambda^{(+,-)}= \chi^{(-1/2)}(x^+) e^{-\frac{i}{2} {x^-}}+ \chi^{(1/2)}(x^+) e^{\frac{i}{2} {x^-}}.
\end{align}
Thus the gauge parameter $\Lambda$ is given in terms of the six chiral functions $\{\lambda^{a}(x^+),\lambda(x^+),\chi^{s}(x^+)\}$ which produce the following variations on the background fields in \eqref{sol_space}:
\begin{align}
	\label{residual-GTs}
	\delta T &=  2 T \, \lambda' + \lambda \, T' +\frac{1}{2}  \lambda''' + i \epsilon^{rs} (3 G_r \chi_s' + G'_r \chi_s) +2 (\lambda^a)^{rs} J_a G_r \chi_s  , \cr
	\delta J_a &= \lambda_a' + i \, {f^{bc}}_a J_b \lambda_c + 4 (\lambda_a)^{rs} G_r \chi_s ,\cr
	\delta G_s &= \lambda G_s' + \frac{3}{2} G_s \lambda' - \chi_s'' - i {(\lambda^a)^r}_s (J_a \, \lambda - \lambda_a) \, G_r - \left( \frac{1}{4} \eta^{ab} J_a J_b + T \right) \chi_s \cr
	& + i {(\lambda^a)^r}_s \left( 2 J_a \, \chi_r'+ J_a' \, \chi_r  \right).
\end{align}
We also have the following expressions
\begin{align}
	{\rm Tr} \left( \Lambda \delta A_- \right) &= \lambda^{(1)} \, (e^{-i x^-}  \delta J^{(-1)} + e^{i x^-} \delta J^{(1)}) , \cr
	{\rm Tr} \left( \Lambda \delta A_+ \right) &= \eta^{ab} \delta J_a \lambda_b - 4i \, \epsilon^{rs} \chi_s \delta G_r -2 \lambda \, \delta T \cr
	& - \partial_+ \partial_- \left(\lambda^{(1)} \, (e^{-i x^-}  \delta J^{(-1)} + e^{i x^-} \delta J^{(1)})\right) .
\end{align}
Using the coordinates $x^\pm  = \phi \pm  t$, we have ${\cal A}_\phi={\cal A}_+ + {\cal A}_-$ and $\partial_\phi = \partial_+ + \partial_-$. The charge that generate residual gauge transformations is \cite{Compere:2018aar}
\begin{align}
	/ \!\!\! \delta Q_\Lambda &= - \frac{k}{2\pi} \int_0^{2\pi} d\phi \, {\rm Tr} \left[ \Lambda \, \delta{\cal A}_\phi \right] \cr
	&= - \frac{k}{2\pi} \int_0^{2\pi} d\phi \, \Big[\eta^{ab} \delta J_a \lambda_b + 4i \, \epsilon^{sr} \chi_s \delta G_r -2 \lambda \, \delta T \cr
	& +  \partial_\phi \left(i \, \lambda^{(1)} \, (e^{-i x^-}  \delta J^{(-1)} - e^{i x^-} \delta J^{(1)}) \right) \Big].
\end{align}
Assuming that $\delta J^{(\pm 1)}(x^+)$ and $\lambda^{(1)}(x^+)$ are periodic in $\phi \rightarrow \phi + 2\pi$ we will drop the total derivative terms here in the second line.
%Thus we see (just as in \cite{aps}) it is integrable and we write
%
%\begin{align}
%Q_\Lambda &=& - \frac{k}{2\pi} \int_0^{2\pi} d\phi \,  \Big[\eta^{ab} J_a \lambda_b - 4i \, \epsilon^{rs} \chi_s \, G_r -2 \lambda \, \, T \cr
%&& ~~~~~~~~~~~~~~~~~~~~~ +  \partial_\phi \left(i \, \lambda^{(1)} \, (e^{-i x^-}  \, J^{(-1)} - e^{i x^-} \, J^{(1)}) \right) \Big]
%\end{align}
%
%Assuming that $J^{(\pm 1)}(x^+)$ and $\lambda^{(1)}(x^+)$ are %periodic in $\phi \rightarrow \phi + 2\pi$ we end up have
%
%\begin{align}
%Q_\Lambda = - \frac{k}{2\pi} \int_0^{2\pi} d\phi \,  \Big[\eta^{ab} J_a \lambda_b - 4i \, \epsilon^{rs} \chi_s \, G_r -2 \lambda \, T \Big]
%\end{align}
%
Before carrying on we will make the following replacements 
\begin{align}
	\lambda_a &\rightarrow  \lambda_a + J_a \, \lambda \cr
	\delta T &\rightarrow \delta T + \frac{1}{2} \eta^{ab} J_a \, \delta J_b \label{modifications}
\end{align}
which modifies the transformations (\ref{residual-GTs}) to
\begin{align}
	\delta T &=  2 T \, \lambda' + \lambda \,  T' +\frac{1}{2}  \lambda''' + i \epsilon^{rs} (3 G_r \chi_s' + G'_r \chi_s) - \frac{1}{2} \eta^{ab} J_a {\lambda}'_b   ,\cr
	\delta J_a &= (\lambda \, J_a)' + \lambda_a' + i \, {f^{bc}}_a J_b \, \lambda_c + 4 (\lambda_a)^{rs} G_r \chi_s ,\cr
	\delta G_s &= \lambda G_s' + \frac{3}{2} G_s \lambda' - \chi_s'' - \left( \frac{1}{2} \eta^{ab} J_a J_b + T \right) \chi_s + i {(\lambda^a)^r}_s \left( 2 J_a \, \chi_r'+ J_a' \, \chi_r +  \lambda_a \, G_r \right) .\cr & \label{mod_trans_first}
\end{align}
and $/ \!\!\! \delta Q_\Lambda$, after replacements \eqref{modifications} remains form-invariant. Thus we see it is integrable and finally we write
\begin{align}
	Q_\Lambda &=  \frac{k}{2\pi} \int_0^{2\pi} \!\!\! d\phi \,  \Big(2 \lambda \, T - \eta^{ab} J_a \lambda_b + 4i \, \epsilon^{rs}  \, G_r \, \chi_s \Big) \, . \label{first_order_charges}
\end{align}
The transformations \eqref{mod_trans_first} and the charge \eqref{first_order_charges} are identical to \eqref{trans_CSG} and \eqref{secondordercharge} from the second order formulation of conformal gravity for  $l=-1$ and $k \rightarrow \frac{k}{16}$ which  results in the identical Poisson bracket and commutation relations derived from the second-order formalism in \eqref{Poissbrac1}, \eqref{Poissbrac2} and \eqref{semi_class_GS3}.
\section{Discussion}
\label{discussion}
We have shown that the 3d conformal gravity admits consistent chiral boundary conditions that lead to an asymptotic symmetry algebra that is the semi-classical limit of the chiral $\Lambda$-$\mathfrak{bms}_4$ algebra \eqref{RESULT} of \cite{Romans:1990ta, Gupta:2022mdt}, one of the four chiral ${\cal W}$-algebra extension of $\mathfrak{so}(2,3)$ \cite{Gupta:2023fmp}. We have done this both in the metric and first-order formalisms of this theory. Our result complements that of \cite{Fuentealba:2020zkf} where the conformal $\mathfrak{bms}_3$, yet another chiral algebra extension of $\mathfrak{so}(2,3)$ is shown to emerge as the ASA of 3d conformal gravity with a different set of boundary conditions in its first-order formalism. It will be interesting  to develop the holography of 3d conformal gravity with these boundary conditions further. 

It is worth noting that, unlike the case of AdS$_3$ gravity the 
configuration space in the 3d conformal gravity, in our gauge appears 
to have more than one pair of conjugate variables: (i) $g^{(0)}_{ab}$ 
and $\epsilon^{ac} g^{(0)}_{cd} g_{(2)}^{db} + \cdots$, as well as 
(ii) $g^{(1)}_{ab}$ and $\epsilon^{ac} g^{(0)}_{cd} g_{(1)}^{db}$ as 
can be read out from \eqref{final_bdry_term_CSG}. In the case of  AdS$_3$ 
gravity, one interprets $g^{(0)}_{ab}$ to be the metric on the 2d 
boundary on which the holographic 2d QFT exists. It is 
important to ask what is the boundary geometry and its symmetries  in the case of 3d conformal gravity. The fact that there exist more than one 
set of (symmetric rank-2 tensor worth of) `coordinates' along with 
their conjugate `momenta' in the phase space suggests that the specification of this 2d 
geometry should include, in addition to $g^{(0)}_{ab}$ some components 
of $g^{(1)}_{ab}$. It will be interesting to explore this enhanced 
geometry and the associated local gauge transformations (going beyond 
the 2d diffoemorphisms and Weyl transformations) of $g^{(0)}_{ab}$ and 
$g^{(1)}_{ab}$ and construct field theories that respect these local 
symmetries coupled to these tensors.

 For locally AdS$_4$ geometries, the induced effective on-shell action goes to zero \cite{Skenderis:1999nb} and in that case addition of Pontryagin term will provide an effective gravitational action \eqref{CSG1} at the boundary that can be used to calculate the expectation value of the symmetry generating currents in the bulk AdS$_4$ gravity. This can also potentially explain the emergence of line integral charges from the bulk AdS$_4$ gravity because the effective action is a 3d gravitational theory which will have its co-dimension two charges making the charges co-dimension three integrals from the bulk perspective. As it turned out the charges we derived from the action \eqref{CSG1} are similar to the conjectured charges from the AdS$_4$ gravity in \cite{Gupta:2022mdt}. We leave further implications of our computations for AdS$_4$/CFT$_3$ for future works.
%		Having dispensed with the motivations, we now turn to the analysis of $3d$ gravitational Chern-Simons action leaving further implications on bulk AdS$_4$ gravity for future works.
%
\section*{Acknowledgements}
We thank the participants of CSM2023 at IMSc for useful comments on the contents of this work. We thank Nirmal Ghorai for useful discussion on gauge that fixes Weyl symmetry. NG would also like to thank Adarsh Sudhakar for useful discussions on covariant phase space formalism. The work of NG is supported by the Grant (File NoSB/SJF/2021-22/14) of the Department of Science and Technology and SERB, India

\appendix
\section{Charges from Lee-Wald covariant phase space formalism}
\label{rev_CSG}
Here we will review the formalism to calculate the charges for Lagrangian density that under symmetry transformations are not covariant and varies by total derivative. We will use the notation and formulae of \cite{Tachikawa:2006sz, Dengiz:2020fpe}. Consider Lagrangian density $L(\phi)$ ($\phi$ denotes the collection of various fields)  and $\chi=\{\xi, \sigma\}$ as the collection of symmetry parameters such that
\begin{align}
	\delta_{\chi}L(\phi)= \mathcal L_{\chi}L(\phi)+\partial_{\mu}\Xi^{\mu}_{(\chi)}, \label{mod_lag}
\end{align}
 where $\mathcal L_{\chi}L$ appears only for diffeomorphisms $\xi$. In the terminology of \cite{Freidel:2021cjp}, the Lagrangian is \textit{covariant} under diffeomorphisms if  $\delta_{\xi}L=\mathcal L_{\xi}L= \partial_{\mu}\left(\xi^{\mu} L\right)$. It is called \textit{semi-covariant} if it satisfies \eqref{mod_lag} for non-zero $\Xi^{\mu}_{(\xi)}$ which is referred to as the Lagrangian anomaly. The Lagrangian density of our theory in \eqref{CSG1} is semi-covariant as its variation with respect to diffeomorphisms in \eqref{diff_var} contains non zero $\Xi^{\mu}_{(\xi)}$  which modifies the analysis of Noether charges.  An arbitrary variation of the Lagrangian density is given by
\begin{align}
	\delta L= \sum_{\phi}\,E_{\phi}\delta \phi+\partial_{\mu} \Theta^{\mu}(\phi,\delta \phi).
\end{align}
$E_{\phi}$ is the equations of motion associated with $\phi$ and $\Theta^{\mu}$ is a presymplectic potential. The on-shell Noether current is then given by
\begin{align}
	J^{\mu}_{\chi}(\phi)=\Theta^{\mu}(\phi, \delta_{\chi}\phi)-\xi^{\mu}L-\Xi^{\mu}_{(\chi)}(\phi), \label{Noethercurrent}
\end{align}
which is conserved $\partial_{\mu}J^{\mu}_{\chi}(\phi)\simeq 0$, where $\simeq$ is the symbol for on-shell equality. Note the modification of Noether current in presence of anomaly. One can construct $K^{\mu\nu}$ from this current such that
\begin{align}
	J^{\mu}_{\chi}(\phi)\simeq \partial_{\nu} K^{\mu\nu}_{\chi}(\phi). \label{two-form}
\end{align}
For semi-covariant Lagrangian the variation of presymplectic potential under symmetry transformation takes the form
\begin{align}
	\delta_{\chi}\Theta^{\mu}(\phi, \delta \phi)= \mathcal L_{\chi} \Theta^{\mu}+\Pi^{\mu}_{\chi}(\phi, \delta \phi),\label{varpresym}
\end{align}
where $\mathcal L_{\chi} \Theta^{\mu}$  once again appears only for the diffeomorphisms. The Lee-Wald symplectic potential for transformation under parameter $\chi$ is given by \cite{Freidel:2021cjp, Campiglia:2020qvc},
\begin{align}
	w^{\mu}(\phi, \delta \phi, \delta_{\chi}\phi)=\frac{1}{16 \pi}\left(\delta \Theta^{\mu}(\phi, \delta_{\chi}\phi)-\delta_{\chi}\Theta^{\mu}(\phi,\delta \phi)-\Theta^{\mu}(\phi, \left[\delta, \delta_{\chi}\right]\phi)\right). \label{LW}
\end{align}
The presence of the last term in \eqref{LW} is to compensate for the background dependence of the symmetry parameter $\chi$ and can be written as follows using \cite{Freidel:2021cjp}, 
\begin{align}
	\Theta^{\mu}\left(\phi, \left[\delta, \delta_{\chi}\right]\phi\right)=\Theta^{\mu}(\phi, \delta_{\delta\chi}\phi)
\end{align}
Using \eqref{Noethercurrent} and \eqref{two-form}, the first term becomes
\begin{align}
	\delta \Theta^{\mu}(\phi,\delta_{\chi}\phi)\simeq \delta \partial_{\nu}K^{\mu \nu}_{\chi}(\phi)+ \xi^{\mu}\delta L(\phi)+\delta \xi^{\mu}\,L(\phi)+ \delta \Xi^{\mu}_{\chi}(\phi) .\label{deltapresym}
\end{align}
The third term is given as
\begin{align}
	\Theta^{\mu}(\phi, \delta_{\delta\chi}\phi)\simeq \partial_{\nu} K^{\mu\nu}_{\delta \chi}(\phi)+ \delta \xi^{\mu}\,L(\phi)+ \Xi^{\mu}_{\delta \chi}(\phi) .\label{presym}   
\end{align}
Substituting \eqref{deltapresym}, \eqref{presym} and \eqref{varpresym} in \eqref{LW} one gets, 
\begin{align}
	&w^{\mu}(\phi, \delta\phi, \delta_{\chi}\phi)\simeq\frac{1}{16 \pi}\,\left(\delta \partial_{\nu}K^{\mu\nu}_{\chi}(\phi)-\partial_{\nu}K^{\mu\nu}_{\delta \chi}+ \delta \Xi^{\mu}_{\chi}(\phi)-\Xi^{\mu}_{\delta \chi}(\phi)-{\Pi}^{\mu}_{\chi}(\phi, \delta\phi)\right) \nonumber \\
	& +\xi^{\mu} \partial_{\rho}\Theta^{\rho} (\phi, \delta \phi)- \mathcal{L}_{\xi}\Theta^{\mu}(\phi, \delta \phi). \label{wd}
\end{align}

The last two terms in \eqref{wd} only exists for the diffeomorphism symmetry parameter.
Using the identity in \cite{Wald:1990mme,Freidel:2021cjp}, one can write
\begin{align}
	\delta \Xi^{\mu}_{\chi}(\phi)-\Xi^{\mu}_{\delta \chi}(\phi)-{\Pi}^{\mu}_{\chi}(\phi, \delta\phi) \simeq \partial_{\nu}\Sigma^{\mu\nu}_{\chi} . \label{totaldiv}
\end{align}
 $\Sigma^{\mu \nu}_{(\chi)}$ is referred to as the symplectic anomaly.
Therefore the final expression of \eqref{LW} after substituting everything is, 
\begin{align}
	w^{\mu}(\phi, \delta \phi, \delta_{\chi}\phi)\simeq\partial_{\nu} Q^{\mu\nu}_{\chi}(\phi, \delta \phi), \label{2-form}
\end{align}
where $Q^{\mu\nu}_{\chi}$ is given by
\begin{align}
	Q^{\mu\nu}_{\chi}=\frac{1}{16 \pi}\left(\delta K^{\mu\nu}_{\chi}(\phi)-K^{\mu\nu}_{\delta \chi}(\phi)+2\,\xi^{[\mu}\Theta^{\nu]}(\phi,\delta\phi)+\Sigma^{\mu\nu}_{\chi}(\phi,\delta \phi)\right). \label{final_two_form}
\end{align}
\eqref{final_two_form} is the final expression for the surface charge for Lagrangian density that transform as a total derivative and are not covariant under the symmetry transformations. It has to be integrated appropriately on a codimension-two hypersurface.

\bibliographystyle{utphys}
%\bibliography{aps01}
\providecommand{\href}[2]{#2}\begingroup\raggedright

\end{document}